\documentstyle[12pt,axodraw]{article}
\begin{document}
\newcommand{\bfig}{\begin{center}\begin{picture}}
\newcommand{\efig}[1]{\end{picture}\\{\small #1}\end{center}}
\newcommand{\flin}[2]{\ArrowLine(#1)(#2)}
\newcommand{\wlin}[2]{\DashLine(#1)(#2){2.5}}
\newcommand{\zlin}[2]{\DashLine(#1)(#2){5}}
\newcommand{\glin}[3]{\Photon(#1)(#2){2}{#3}}
\newcommand{\lin}[2]{\Line(#1)(#2)}
\newcommand{\sof}{\SetOffset}
\newcommand{\bmip}[2]{\begin{minipage}[t]{#1pt}\bfig(#1,#2)}
\newcommand{\emip}[1]{\efig{#1}\end{minipage}}
\newcommand{\putk}[2]{\Text(#1)[r]{$p_{#2}$}}
\newcommand{\putp}[2]{\Text(#1)[l]{$p_{#2}$}}
\newcommand{\bq}{\begin{equation}}
\newcommand{\eq}{\end{equation}}
\newcommand{\bqa}{\begin{eqnarray}}
\newcommand{\eqa}{\end{eqnarray}}
\newcommand{\nl}{\nonumber \\}
\newcommand{\eqn}[1]{eq. (\ref{#1})}
\newcommand{\ibidem}{{\it ibidem\/},}
\newcommand{\into}{\;\;\to\;\;}
\newcommand{\epl}{e^+}
\newcommand{\emn}{e^-}
\newcommand{\nue}{\nu_e}
\newcommand{\nueb}{\bar{\nu}_e}
\newcommand{\mpl}{\mu^+}
\newcommand{\mmn}{\mu^-}
\newcommand{\num}{\nu_{\mu}}
\newcommand{\numb}{\bar{\nu}_{\mu}}
\newcommand{\tpl}{\tau^+}
\newcommand{\tmn}{\tau^-}
\newcommand{\nut}{\nu_{\tau}}
\newcommand{\nutb}{\bar{\nu}_{\tau}}
\newcommand{\ubar}{\bar{u}}
\newcommand{\dbar}{\bar{d}}
\newcommand{\cbar}{\bar{c}}
\newcommand{\sbar}{\bar{s}}
\newcommand{\bbar}{\bar{b}}
\newcommand{\ww}[2]{\langle #1 #2\rangle}
\newcommand{\wws}[2]{\langle #1 #2\rangle^{\star}}
\newcommand{\smod}{\tilde{\sigma}}
\newcommand{\dilog}[1]{\mbox{Li}_2\left(#1\right)}
\newcommand{\umu}{^{\mu}}
\newcommand{\cjg}{^{\star}}
\newcommand{\lgn}[1]{\log\left(#1\right)}
\newcommand{\si}{\sigma}
\newcommand{\sit}{\sigma_{tot}}
\newcommand{\sqs}{\sqrt{s}}
\newcommand{\sih}{\hat{\sigma}}
\newcommand{\sith}{\hat{\sigma}_{tot}}
\newcommand{\p}[1]{{\scriptstyle{\,(#1)}}}
\newcommand{\res}[3]{$#1 \pm #2~~\,10^{-#3}$}
\newcommand{\rrs}[2]{\multicolumn{1}{l|}{$~~~.#1~~10^{#2}$}}
\newcommand{\err}[1]{\multicolumn{1}{l|}{$~~~.#1$}}
\newcommand{\ru}[1]{\raisebox{-.2ex}{#1}}
\title{
\vspace{-4cm}
\begin{flushright}
{\large  PSI-PR-96-10} \\
{\large  FERMILAB-PUB-96/057-T}
\end{flushright}
\vspace{4cm}
Final state QCD corrections to off-shell single top production in
hadron collisions.}
\author{R.~Pittau\thanks{email address: pittau@psw218.psi.ch}\\
        Paul Scherrer Institute, CH-5232 Villigen-PSI, Switzerland }
\maketitle
\thispagestyle{empty}
\begin{abstract}
In this paper, I study final state 
QCD radiative corrections to off-shell single top production via
decaying $W$ at hadron collider energies. 
The tree level Breit-Wigner distribution of 
the produced top invariant mass is distorted by final state QCD radiation, 
while the peak position remains unchanged.  
The exact one loop QCD calculation 
and the narrow width approximation agree in predicting 
the cross section, the hadronic transverse 
energy distribution and the bottom-lepton invariant mass distribution.
\end{abstract}
\clearpage
\setcounter{page}{1}
\section{Introduction}
The discovery of the top quark at CDF and D0 \cite{cit1}, has
opened a new era of measurements in top-quark physics.
Now, the properties of the top quark can be directly investigated, not only
inferred from their effects in radiative corrections.

\noindent At hadron colliders, the dominant production mechanisms
are, of course, the $t \bar t$ channels
\bqa
& &q\, \bar{q}\, \to \,t\, \bar{t} \nonumber \\
& &g\, g\, \to \,t\, \bar t~, 
\label{dotop}
\eqa
but single top quarks events are also present, such as
\bqa
& &q^{\prime}\, g\, (W^{+}\, g)\, \to\, q\, t\, \bar{b}\nonumber \\  
& &q^{\prime}\, b\,\to\, q\, t \nonumber \\
& &q^{\prime}\, \bar q\, \to\, W^{*}\, \to\, t\, \bar b \nonumber \\
& &g\, b\, \to\, W^{-}\, t ~.
\label{sitop}
\eqa
The first two mechanisms are known as W-gluon processes \cite{cit3},
the third one as $W^*$ production \cite{cit4} and the fourth one as $W t$ 
production \cite{cit5}. The cross sections in \eqn{sitop} are ordered 
according to their magnitude in $p \,\bar p$ collisions at $\sqrt{s}=~2$ TeV
for $m_t=~176$ GeV \cite{cit7}.

\noindent Even with less expected events, single top production processes are 
important because they provide a consistency check 
on the measured parameters of the top quark in $t \bar t$ production.

Radiative corrections to the processes in \eqn{dotop} and \eqn{sitop}
are well known in the literature \cite{cit8}, but usually, performed in 
the narrow width approximation,
in which production and decay of the top are treated independently.
A check on the validity of this approach is still missing.
Since, in the narrow width approximation, diagrams connecting 
decaying 
products with the production process are missing, one especially expects 
deviations due to a non exact treatment of the gluonic radiation, 
which is an important quantity for the reconstruction of $m_t$ in $t\, \bar t$ 
events \cite{cit9}. 
Therefore a precise study of it, even in a simpler case, can give hints on 
its relevance in the main production mechanisms of \eqn{dotop}. 

For those reasons, I decided to perform a complete calculation of the final
state QCD radiative corrections to $W^*$ single top production, taking
into account all the subsequent decays. 
\noindent Among all the others, the $W^*$ mechanism is interesting
because possible new physics may introduce a high mass state (say particle $V$)
to couple strongly with the $\bar t b$ system such that the production rate
from $q^{\prime}\, \bar q\, \to\, V\, \to \, t\, \bar b$
can deviate from the standard model $W^*$ rate \cite{cit7,cit10}.
Therefore accurate predictions of the standard properties of the top 
in this channel are also important by themselves.

\noindent The background QCD contribution is known \cite{cit4}, so I shall 
study here the QCD one loop corrections to the signal diagram of fig. 1.
For this process, thanks to the color structure, initial and final state 
QCD corrections do not interfere and are separately gauge invariant, so
that, in order not to obscure the effects I want to study, I decided, 
in this paper, to concentrate my attention on final state gluonic corrections.
Initial state corrections are however very simple and a study 
including them will be treated elsewhere \cite{1}.
I chose the semi leptonic final state $\nu_l~l^+~b~\bar b$
because it is easier to detect experimentally. The extension to 
$q^{\prime}~\bar q~b~\bar b$ is anyway trivial. 
In fact, diagrams with gluons connecting $b$ or $\bar b$ with $q^{\prime}$ 
or $\bar q$ are killed, at the one loop level, by the color factor, so one 
is left with simple gluonic corrections for the $W\,q^\prime\,\bar q$
vertex.    
\bfig(300,130)
\SetScale{1}
\sof(50,20)
\flin{40,0}{65,25} \flin{65,25}{40,50}
\wlin{65,25}{100,25} \flin{125,0}{100,25} \flin{100,25}{125,50}
\wlin{117.5,42.5}{125,75}\flin{125,75}{150,100}\flin{150,50}{125,75}
\Text(35,0)[rt]{$u$}
\Text(35,50)[rb]{$\bar d$}
\Text(130,0)[lt]{$\bar b$}
\Text(130,50)[lb]{$b$}
\Text(155,100)[lb]{$\nu_l$}
\Text(155,50)[lt]{$l^+$}
\Text(108,38)[rb]{$t$}
\efig{Figure 1: {\em
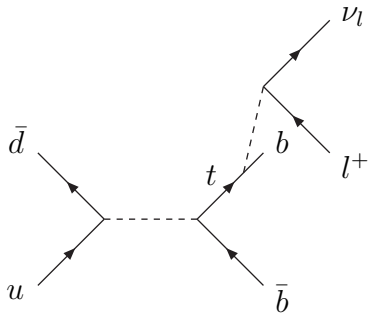
Tree level diagram for $\bar d~u \to \nu_l~l^+~b~\bar b$ via single top 
production. Here, and in the following figures, dashed lines denote $W$'s.}}
\section{The calculation}
The tree level diagram for the process is drawn in fig.1,
while in fig. 2 and 3 I show the one loop virtual diagrams and the real 
bremsstrahlung. 
\bfig(300,230)
\SetScale{1}
\sof(-100,130)
\flin{60,0}{85,25} \flin{85,25}{60,50}
\wlin{85,25}{100,25} \flin{125,0}{100,25} \flin{100,25}{125,50}
\wlin{105,30}{125,75}\flin{125,75}{150,100}\flin{150,50}{125,75}
\GlueArc(116,41)(8.5,-135,45){2}{4}
\sof(40,130)
\flin{60,0}{85,25} \flin{85,25}{60,50}
\wlin{85,25}{100,25} \flin{125,0}{100,25} \flin{100,25}{125,50}
\wlin{115,40}{125,75}\flin{125,75}{150,100}\flin{150,50}{125,75}
\GlueArc(112.5,12.5)(10,-45,135){2}{5}
\sof(180,130)
\flin{60,0}{85,25} \flin{85,25}{60,50}
\wlin{85,25}{100,25} \flin{125,0}{100,25} \flin{100,25}{125,50}
\wlin{119,44}{125,75}\flin{125,75}{150,100}\flin{150,50}{125,75}
\Gluon(110,15)(110,35){-2}{4}
\sof(-100,30)
\flin{60,0}{85,25} \flin{85,25}{60,50}
\wlin{85,25}{100,25} \flin{125,0}{100,25} \flin{100,25}{125,50}
\wlin{115,40}{125,75}\flin{125,75}{150,100}\flin{150,50}{125,75}
\Gluon(120,5)(120,45){-2}{7}
\sof(40,30)
\flin{60,0}{85,25} \flin{85,25}{60,50}
\wlin{85,25}{100,25} \flin{125,0}{100,25} \flin{100,25}{125,50}
\wlin{115,40}{125,75}\flin{125,75}{150,100}\flin{150,50}{125,75}
\GlueArc(112.5,37.5)(10,-135,45){2}{5}
\sof(180,30)
\flin{60,0}{85,25} \flin{85,25}{60,50}
\wlin{85,25}{100,25} \flin{125,0}{100,25} \flin{100,25}{125,50}
\wlin{120,45}{125,75}\flin{125,75}{150,100}\flin{150,50}{125,75}
\GlueArc(110,35)(10,-135,45){2}{5}
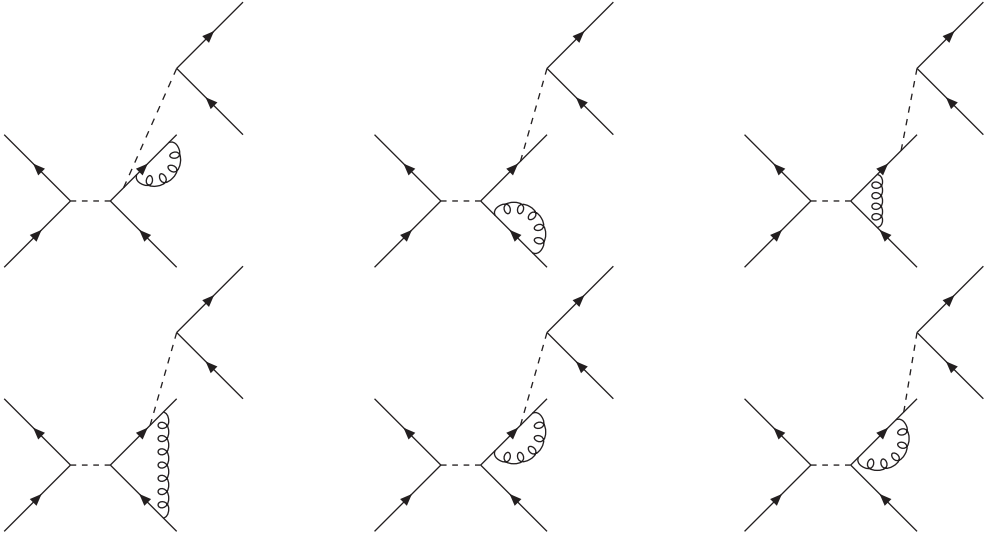
\efig{Figure 2: {\em Final state one loop QCD virtual diagrams.}}
\bfig(300,150)
\SetScale{1}
\sof(-100,30)
\flin{60,0}{85,25} \flin{85,25}{60,50}
\wlin{85,25}{100,25} \flin{125,0}{100,25} \flin{100,25}{125,50}
\wlin{111,36}{125,75}\flin{125,75}{150,100}\flin{150,50}{125,75}
\Gluon(118,43)(150,43){2}{5}
\sof(40,30)
\flin{60,0}{85,25} \flin{85,25}{60,50}
\wlin{85,25}{100,25} \flin{125,0}{100,25} \flin{100,25}{125,50}
\wlin{119,44}{125,75}\flin{125,75}{150,100}\flin{150,50}{125,75}
\Gluon(109,34)(141,34){2}{5}
\sof(180,30)
\flin{60,0}{85,25} \flin{85,25}{60,50}
\wlin{85,25}{100,25} \flin{125,0}{100,25} \flin{100,25}{125,50}
\wlin{119,44}{125,75}\flin{125,75}{150,100}\flin{150,50}{125,75}
\Gluon(110,15)(142,15){2}{5}
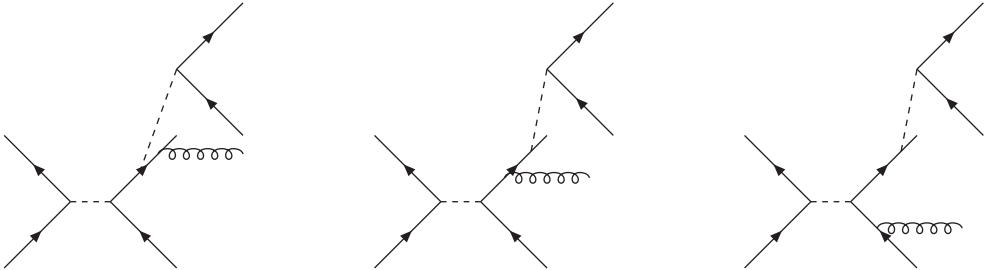
\efig{Figure 3: {\em Real gluon emission.}}
I computed the virtual corrections using standard Passarino-Veltman 
techniques \cite{2}, with the help of the Symbolic Manipulation program
FORM \cite{form}. I used dimensional regularization for
ultraviolet, collinear and soft divergences.
Furthermore, I kept everywhere complex masses for top and $W$, but
I systematically neglected the bottom mass.

\noindent An analytic approximation in $n$ dimensions for the soft-collinear 
part of the real emission was built following ref. \cite{3} and the 
cancellation of all divergences performed analytically.

\noindent Both the virtual and the real contributions have been computed 
applying helicity amplitudes methods \cite{4}, and the final expressions
implemented in a Monte Carlo program \cite{1}, that uses the self-optimization 
techniques of ref. \cite{6}.
Since those techniques are applied here, for the first time, in loop
calculations, it may be useful to briefly discuss the adopted strategy.
More details will be found in ref. \cite{1}.

The problem here is the matching between hard and soft phase space.
Schematically, the final result for the cross section $\sit$
(with any kind of cuts) can be written as a sum of four contributions
\bq
\sit= \si_{0}+\si_{V}+\si_{S}(\delta)+\si_{H}(\delta) 
\label{sit} 
\eq
where $\si_{0}$ is the lowest order result, $\si_{V}$ the virtual
contribution and $\si_{S}$, $\si_{H}$ the soft and hard real
radiation.

\noindent The sum, $\si_{V}+\si_{S}(\delta)$ does not contain soft and collinear
singularities. On the other hand $\si_{S}(\delta)+\si_{H}(\delta)$ 
is independent on $\delta$, where $\delta$ is the separation between soft and 
hard gluons (following  again ref. \cite{3}, $\delta$ in a cut on 
the invariant mass of $g~+~b$ and $g~+~\bar b$). 
The last statement is true only if an {\em exact} 
computation of $\si_{S}(\delta)$ is performed. Instead, what one usually does 
is computing $\si_{S}(\delta)$ for small $\delta$. In such a limit,
because of factorization properties (\cite{3,12}), 
very simple expressions are obtainable in terms of the born result multiplied
by universal coefficients containing $\log(\delta)$ and
$\log^2(\delta)$. At this point, by numerically going to the limit 
$\delta \to 0$, one gets unbiased results. Of course, if $\delta$ is too
small, large numerical cancellations take place between 
$\si_{S}(\delta)$ and $\si_{H}(\delta)$, resulting in large errors in the Monte 
Carlo integration. A good value for $\delta$ can be usually found by 
numerically checking the independence on $\delta$ of the results.

\noindent For fixed $\delta$ one would like to know how many Monte Carlo 
points have to be spent to separately compute all four contributions 
in \eqn{sit}, mainly because usually the most time consuming part is 
$\si_{V}$, that contains
loop diagrams. This is a typical problem that can be solved using the 
Multichannel self-optimizing approach of ref. \cite{6}. One starts with the same
amount of points for all channels and, during the run, the Monte Carlo
self-adjusts itself, so that afterwards one usually obtains  
a smaller percentage of the computational time spent in the
evaluation of $\si_{V}$, which means a better estimate of $\sit$
in a shorter time.

\noindent In table 1 I show a typical result of the self-optimization
procedure.
\begin{table}[h]
\begin{center}
\begin{tabular}{|c|c|c|} \hline\hline
 channel   & percentage before opt. & percentage after opt.\\
\hline
1                 & 0.2 & 0.0996\\
2                 & 0.2 & 0.5436\\
3                 & 0.2 & 0.1265\\
4                 & 0.2 & 0.2083\\
5                 & 0.2 & 0.0220\\
\hline \hline  
\end{tabular}
\end{center}
\caption[.]{{\em Percentage of the Monte Carlo points used for each channel 
in the evaluation of $\sit$ before and after the self-optimization.
The first three channels take care of the
peaking structure of $\si_{H}$ given by the Feynman diagrams in fig.
3, channel 4 refers to $\si_{0}$ + $\si_{S}$
and channel 5 to $\si_{V}$.}} 
\end{table}

I made several checks on the final result. 
First of all, I verified that the $CP$ invariance of the
tree level current
\bq
 T_{\alpha\mu}= \bar{u}(b)\,\gamma_\alpha (1-\gamma_5) 
(\rlap /p_{l^+} + \rlap /p_{\nu_l} + \rlap /p_b+m_t)\gamma_\mu(1-\gamma_5) 
\,v(\bar b) 
\eq
remains after QCD loop corrections.
Then, by numerically rescaling $\Gamma_W$, $\Gamma_t$ and the cross section
by the same amount, I checked the agreement between the Monte
Carlo estimate of the total ${\cal O} (\alpha_s)$  $t \bar b$ on-shell 
cross section
($\si_{MC}$) and (for example) the analytic result ($\si_{AN}$) of
ref. \cite{11} (see table 2).
I also tested, for small $\delta$, the independence on $\delta$ 
of the results.
All numbers in this paper are obtained with
$\delta= 0.2\, GeV^2$. 
\begin{table}[h]
\begin{center}
\begin{tabular}{|l|c|c|} \hline\hline
$\sqrt s$ (GeV)    & $\si_{MC}$ (pb)& $\si_{AN}$\\
\hline
300                &0.09086 $\pm$ 0.00002 &0.09088\\
                   &0.10499 $\pm$ 0.00035 &0.10547\\
\hline
600                &0.03508 $\pm$ 0.00001 &0.03507\\
                   &0.03784 $\pm$ 0.00027 &0.03774\\
\hline
900                &0.01653 $\pm$ 0.00001 &0.01653\\
                   &0.01763 $\pm$ 0.00019 &0.01748\\
\hline
1200               &0.00948 $\pm$ 0.00001 &0.00948\\
                   &0.00999 $\pm$ 0.00014 &0.00996\\
\hline
1500               &0.00612 $\pm$ 0.00001 &0.00612\\
                   &0.00643 $\pm$ 0.00010 &0.00640\\
\hline
1800               &0.00427 $\pm$ 0.00001 &0.00427\\
                   &0.00454 $\pm$ 0.00007 &0.00446\\
\hline \hline  
\end{tabular}
\end{center}
\caption[.]{{\em Comparison between the Monte Carlo total cross section, 
in the limit of vanishing widths, and the analytic on-shell calculation.
No convolution with the parton densities has been performed.
The first entry is the tree level result. In the second entry
all final state QCD corrections are included. 
$m_t=~176\,GeV$ and $\alpha_s=$ 0.12.}}
\end{table}

A last comment is in order.
Taking into account the widths of the decaying particles gives rise
to conceptual problems with respect to the gauge invariance.
The correct gauge invariant prescription would be to compute the
widths as the imaginary part of the one loop renormalized propagators
and all set of loop diagrams necessary to restore gauge invariance.
Since, in the process at hand, $W$ and $t$ decay via electroweak interactions,
this would imply to include terms of the one loop ${\cal O} (\alpha)$ 
calculation at the tree level and part of the two loop 
${\cal O} (\alpha \alpha_s)$ corrections in the ${\cal O} (\alpha_s)$ 
contribution.
Since I am interested here in ${\cal O}(\alpha_s)$ corrections, 
the neglected ${\cal O}(\alpha)$ terms are expected to be small,
so I followed the naive prescription of considering 
everywhere constant complex masses.
For the sake of consistency, when computing $Im(m^2_t)$, 
I used the lowest order top width 
value $\Gamma_t= 1.6429~GeV$ for quantities
at the tree level and the QCD corrected value $\Gamma_t= 1.5117~GeV$ 
when including QCD corrections.
\section{Results}
In this section, I present some results for the process
$\bar d~u \to \nu_l~l^+~b~ \bar b$ obtained with the Monte Carlo
of ref. \cite{1}. I chose to plot three useful distributions 
for measuring the top mass in $p \bar p$ collisions at 
$\sqrt{s}= 2~TeV$, namely the total
hadronic transverse energy ($H_T$), the invariant mass 
$\sqrt{(p_{l^+} + p_{b})^2}$ ($m_{bl}$) and the "top mass distribution" 
$\sqrt{(p_{\nu_l}+p_{l^+}+p_b)^2}$ ($m_{bl\nu}$). Of course, due to the 
presence of an undetected neutrino, the last quantity is not going to be easy 
to reconstruct experimentally. However, $m_{bl\nu}$ is of theoretical 
interest and directly measurable in the channel  
$q^{\prime}~\bar q~b~\bar b$. 

\noindent I used the following input parameters and cuts
\bqa                    
&&\alpha= 1/128,~~\sin^2_{\theta}= 0.2224,~~\alpha_s= 0.1 \nonumber \\
&&M_W= 80.41~GeV,~~M_t= 176~GeV             \nonumber \\
&&\Gamma_W= 2.1185~GeV~~\nonumber \\
&&E_T(\nu_l),~E_T(l^+),~E_T(b),~E_T(\bar b)~
>~15~GeV \nonumber \\
&&|\eta (l^+)|,~|\eta (b)|,~|\eta (\bar b)|~<~2,~~\Delta R (b \bar b) > 0.7~,
\label{cuts}
\eqa
together with the cone jet-definition algorithm of ref. \cite{jetden}
(with jet cone size $R$= 0.7) and the CTEQ2M parton densities of
 ref. \cite{cteq}.
Two jets with $b$ content are required to be present in the visible region 
defined by the above cuts, and no extra (gluonic) jets.
In order to compare the full QCD calculation with the narrow width 
approximation, I also produced histograms for $H_T$ and $m_{bl}$, 
in which $W$ and top are put on-shell by numerically rescaling 
$\Gamma_{W,\,t}$ and the cross section by the same amount.

 In fig. 4 and 5, I show $H_T$ and $m_{bl}$ in the off-shell case and 
in the on-shell limit, including all final state gluonic corrections.
For comparisons, I also show the tree level result.
As one can see, the on-shell and off-shell distributions are almost 
indistinguishable, therefore the narrow width approximation works very well.

\noindent In fig 6. the off-shell top invariant mass distribution 
is shown with and without final state QCD corrections. 
QCD radiation is responsible for a distortion in the Breit-Wigner distribution:
more events are produced in the left tail and less in the right side, but the
position of the peak is essentially unchanged
\footnote{By looking at fig. 4 and 5 one can recognize a similar
distortion in $H_T$ and $m_{bl}$ as well.}. 
Comparing with the narrow width approximation 
is difficult for $m_{bl\nu}$. One would be forced to 
impose a Breit-Wigner by hand.
I did not try that. Fig. 6 already gives a quantitative prediction for
the distortion of the one loop distribution with respect to the tree
level result. A precise quantitative knowledge of this effect may be useful 
in fitting the experimental distributions, and estimating the systematic 
errors.  

\noindent In on-shell production, the cross section (with the cuts and input
parameters of \eqn{cuts}, including final state QCD corrections) 
is $0.02677 \pm 0.00025~pb$, while one gets $0.02681 \pm 0.00024~pb$ in the 
off-shell case.
That means that the narrow width approach can be safely used also in
computing the cross section. The question of the
total number of produced events is important
when looking at the single top production rate in this channel for new 
physics searches.

\section{Conclusions}
I have performed a complete one loop calculation
of the final state QCD corrections to the single top production process 
$\bar d~u \to \nu_l~l^+~b~ \bar b$ in the off-shell case.
Final state gluonic radiation is responsible for
a distortion in the distributions useful for top mass reconstruction.
The validity of the narrow width approximation is confirmed at the
level of accuracy one naively expects, namely ${\cal O}(\Gamma_t/m_t)$.
However, one should observe that, no one loop QCD diagrams connecting initial 
and final states can contribute to the process at hand. On the other hand, 
such diagrams are present in the main production mechanisms of \eqn{dotop}. 
Therefore, in order to check
the narrow width approach in that more general case, a full off-shell 
QCD loop calculation for the $t \bar t$ channels is still needed.
\section*{Acknowledgments}
I acknowledge W. Giele for stimulating discussions. This work was
partly written during my visit at FERMILAB theory group, that I 
thank for the warm hospitality.
I also wish to thank F. Cuypers for reading the manuscript 
of this paper. 

\clearpage
\begin{center} 
  \begin{picture}(320,380)
\SetOffset(10,30)
  \LinAxis(30,0)(300,0)(9,3,5,0,1.5)
  \LinAxis(30,300)(300,300)(9,3,-5,0,1.5)
  \LinAxis(30,0)(30,300)(6,5,-5,0,1.5)
  \LinAxis(300,0)(300,300)(6,5,5,0,1.5)
  \Text(60,-10)[t]{$60$}
  \Text(120,-10)[t]{$120$}
  \Text(180,-10)[t]{$180$}
  \Text(240,-10)[t]{$240$}
  \Text(280,-10)[t]{{\boldmath $[GeV]$}}
  \Text(10,50)[r]{$5 \cdot 10^{-4}$}
  \Text(10,100)[r]{$10 \cdot 10^{-4}$}
  \Text(10,150)[r]{$15 \cdot 10^{-4}$}
  \Text(10,200)[r]{$20 \cdot 10^{-4}$}
  \Text(10,250)[r]{$25 \cdot10^{-4}$}
  \Text(10,280)[r]{{\boldmath $[\frac{pb}{10~GeV}]$}}
\Line( 30.000,  2.735)( 40.000,  2.735)
\Line( 40.000,  2.735)( 40.000, 13.506)
\Line( 40.000, 13.506)( 50.000, 13.506)
\Line( 50.000, 13.506)( 50.000, 36.615)
\Line( 50.000, 36.615)( 60.000, 36.615)
\Line( 60.000, 36.615)( 60.000, 70.338)
\Line( 60.000, 70.338)( 70.000, 70.338)
\Line( 70.000, 70.338)( 70.000,138.415)
\Line( 70.000,138.415)( 80.000,138.415)
\Line( 80.000,138.415)( 80.000,231.846)
\Line( 80.000,231.846)( 90.000,231.846)
\Line( 90.000,231.846)( 90.000,270.462)
\Line( 90.000,270.462)(100.000,270.462)
\Line(100.000,270.462)(100.000,252.923)
\Line(100.000,252.923)(110.000,252.923)
\Line(110.000,252.923)(110.000,247.538)
\Line(110.000,247.538)(120.000,247.538)
\Line(120.000,247.538)(120.000,215.385)
\Line(120.000,215.385)(130.000,215.385)
\Line(130.000,215.385)(130.000,186.000)
\Line(130.000,186.000)(140.000,186.000)
\Line(140.000,186.000)(140.000,159.077)
\Line(140.000,159.077)(150.000,159.077)
\Line(150.000,159.077)(150.000,133.138)
\Line(150.000,133.138)(160.000,133.138)
\Line(160.000,133.138)(160.000,113.831)
\Line(160.000,113.831)(170.000,113.831)
\Line(170.000,113.831)(170.000, 92.477)
\Line(170.000, 92.477)(180.000, 92.477)
\Line(180.000, 92.477)(180.000, 78.215)
\Line(180.000, 78.215)(190.000, 78.215)
\Line(190.000, 78.215)(190.000, 64.077)
\Line(190.000, 64.077)(200.000, 64.077)
\Line(200.000, 64.077)(200.000, 55.185)
\Line(200.000, 55.185)(210.000, 55.185)
\Line(210.000, 55.185)(210.000, 47.338)
\Line(210.000, 47.338)(220.000, 47.338)
\Line(220.000, 47.338)(220.000, 39.262)
\Line(220.000, 39.262)(230.000, 39.262)
\Line(230.000, 39.262)(230.000, 36.462)
\Line(230.000, 36.462)(240.000, 36.462)
\Line(240.000, 36.462)(240.000, 29.462)
\Line(240.000, 29.462)(250.000, 29.462)
\Line(250.000, 29.462)(250.000, 23.477)
\Line(250.000, 23.477)(260.000, 23.477)
\Line(260.000, 23.477)(260.000, 20.554)
\Line(260.000, 20.554)(270.000, 20.554)
\Line(270.000, 20.554)(270.000, 17.062)
\Line(270.000, 17.062)(280.000, 17.062)
\Line(280.000, 17.062)(280.000, 17.123)
\Line(280.000, 17.123)(290.000, 17.123)
\Line(290.000, 17.123)(290.000, 12.234)
\Line(290.000, 12.234)(300.000, 12.234)
\Line(300.000, 12.234)(300.000,  0.000)
\DashLine( 30.000,  2.683)( 40.000,  2.683){2}
\DashLine( 40.000,  2.683)( 40.000, 12.644){2}
\DashLine( 40.000, 12.644)( 50.000, 12.644){2}
\DashLine( 50.000, 12.644)( 50.000, 32.948){2}
\DashLine( 50.000, 32.948)( 60.000, 32.948){2}
\DashLine( 60.000, 32.948)( 60.000, 68.280){2}
\DashLine( 60.000, 68.280)( 70.000, 68.280){2}
\DashLine( 70.000, 68.280)( 70.000,129.560){2}
\DashLine( 70.000,129.560)( 80.000,129.560){2}
\DashLine( 80.000,129.560)( 80.000,232.800){2}
\DashLine( 80.000,232.800)( 90.000,232.800){2}
\DashLine( 90.000,232.800)( 90.000,268.320){2}
\DashLine( 90.000,268.320)(100.000,268.320){2}
\DashLine(100.000,268.320)(100.000,258.720){2}
\DashLine(100.000,258.720)(110.000,258.720){2}
\DashLine(110.000,258.720)(110.000,244.280){2}
\DashLine(110.000,244.280)(120.000,244.280){2}
\DashLine(120.000,244.280)(120.000,207.920){2}
\DashLine(120.000,207.920)(130.000,207.920){2}
\DashLine(130.000,207.920)(130.000,185.240){2}
\DashLine(130.000,185.240)(140.000,185.240){2}
\DashLine(140.000,185.240)(140.000,155.960){2}
\DashLine(140.000,155.960)(150.000,155.960){2}
\DashLine(150.000,155.960)(150.000,137.120){2}
\DashLine(150.000,137.120)(160.000,137.120){2}
\DashLine(160.000,137.120)(160.000,117.200){2}
\DashLine(160.000,117.200)(170.000,117.200){2}
\DashLine(170.000,117.200)(170.000, 97.640){2}
\DashLine(170.000, 97.640)(180.000, 97.640){2}
\DashLine(180.000, 97.640)(180.000, 85.600){2}
\DashLine(180.000, 85.600)(190.000, 85.600){2}
\DashLine(190.000, 85.600)(190.000, 68.200){2}
\DashLine(190.000, 68.200)(200.000, 68.200){2}
\DashLine(200.000, 68.200)(200.000, 53.560){2}
\DashLine(200.000, 53.560)(210.000, 53.560){2}
\DashLine(210.000, 53.560)(210.000, 50.560){2}
\DashLine(210.000, 50.560)(220.000, 50.560){2}
\DashLine(220.000, 50.560)(220.000, 42.920){2}
\DashLine(220.000, 42.920)(230.000, 42.920){2}
\DashLine(230.000, 42.920)(230.000, 35.016){2}
\DashLine(230.000, 35.016)(240.000, 35.016){2}
\DashLine(240.000, 35.016)(240.000, 29.820){2}
\DashLine(240.000, 29.820)(250.000, 29.820){2}
\DashLine(250.000, 29.820)(250.000, 24.888){2}
\DashLine(250.000, 24.888)(260.000, 24.888){2}
\DashLine(260.000, 24.888)(260.000, 20.116){2}
\DashLine(260.000, 20.116)(270.000, 20.116){2}
\DashLine(270.000, 20.116)(270.000, 17.504){2}
\DashLine(270.000, 17.504)(280.000, 17.504){2}
\DashLine(280.000, 17.504)(280.000, 15.556){2}
\DashLine(280.000, 15.556)(290.000, 15.556){2}
\DashLine(290.000, 15.556)(290.000, 12.928){2}
\DashLine(290.000, 12.928)(300.000, 12.928){2}
\DashLine(300.000, 12.928)(300.000,  0.000){2}
\DashCurve{( 40.000,  1.758)
( 50.000,  9.520)
( 60.000, 25.800)
( 70.000, 54.480)
( 80.000,104.080)
( 90.000,180.240)
(100.000,215.920)
(110.000,212.320)
(120.000,200.200)
(130.000,185.400)
(140.000,165.720)
(150.000,146.280)
(160.000,127.760)
(170.000,110.320)
(180.000, 94.920)
(190.000, 81.440)
(200.000, 69.880)
(210.000, 59.320)
(220.000, 50.440)
(230.000, 42.920)
(240.000, 36.600)
(250.000, 31.236)
(260.000, 26.652)
(270.000, 22.812)
(280.000, 19.520)
(290.000, 16.616)
(300.000, 14.160)}{0.9}    
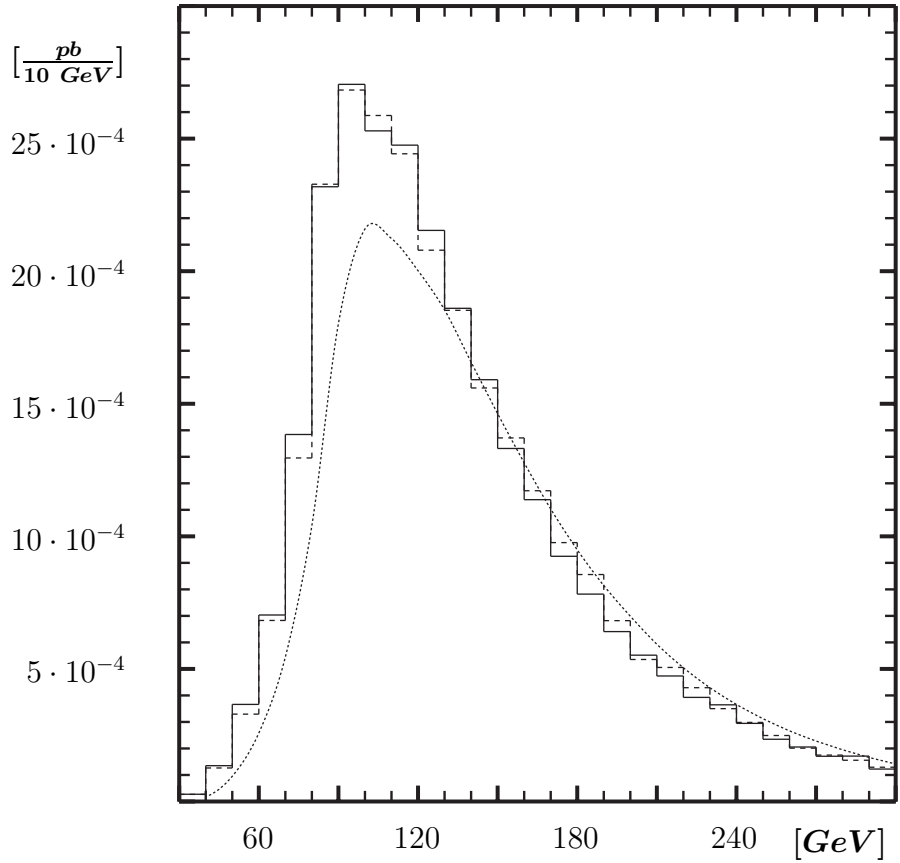
\efig{Figure 4: {\em Total hadronic transverse energy for
on-shell (dashed histogram) and off-shell (solid histogram) single
top production. Final state QCD corrections included.
The dotted line is the off-shell tree level result ($\alpha_s= 0$).}}
  \begin{center}                               
  \begin{picture}(340,330)                     
\SetOffset(30,30)
  \LinAxis(0,0)(300,0)(15,2,5,0,1.5)           
  \LinAxis(0,200)(300,200)(15,2,-5,0,1.5)      
  \LinAxis(0,0)(0,200)(5,4,-5,0,1.5)           
  \LinAxis(300,0)(300,200)(5,4,5,0,1.5)        
  \Text(60,-10)[t]{$30$}                       
  \Text(120,-10)[t]{$60$}                      
  \Text(180,-10)[t]{$90$}                      
  \Text(240,-10)[t]{$120$}                     
  \Text(280,-10)[t]{{\boldmath $[GeV]$}}
  \Text(-20,40)[r]{$4 \cdot 10^{-4}$}
  \Text(-20,80)[r]{$8 \cdot 10^{-4}$}
  \Text(-20,120)[r]{$12\cdot 10^{-4}$}
  \Text(-20,160)[r]{$16 \cdot 10^{-4}$}
  \Text(-20,187)[r]{{\boldmath $[\frac{pb}{5~GeV}]$}}
\Line(  0.000,  2.869)( 10.000,  2.869)
\Line( 10.000,  2.869)( 10.000,  7.503)
\Line( 10.000,  7.503)( 20.000,  7.503)
\Line( 20.000,  7.503)( 20.000,  8.794)
\Line( 20.000,  8.794)( 30.000,  8.794)
\Line( 30.000,  8.794)( 30.000, 12.514)
\Line( 30.000, 12.514)( 40.000, 12.514)
\Line( 40.000, 12.514)( 40.000, 19.831)
\Line( 40.000, 19.831)( 50.000, 19.831)
\Line( 50.000, 19.831)( 50.000, 22.292)
\Line( 50.000, 22.292)( 60.000, 22.292)
\Line( 60.000, 22.292)( 60.000, 29.323)
\Line( 60.000, 29.323)( 70.000, 29.323)
\Line( 70.000, 29.323)( 70.000, 36.246)
\Line( 70.000, 36.246)( 80.000, 36.246)
\Line( 80.000, 36.246)( 80.000, 42.785)
\Line( 80.000, 42.785)( 90.000, 42.785)
\Line( 90.000, 42.785)( 90.000, 52.985)
\Line( 90.000, 52.985)(100.000, 52.985)
\Line(100.000, 52.985)(100.000, 65.892)
\Line(100.000, 65.892)(110.000, 65.892)
\Line(110.000, 65.892)(110.000, 71.354)
\Line(110.000, 71.354)(120.000, 71.354)
\Line(120.000, 71.354)(120.000, 78.092)
\Line(120.000, 78.092)(130.000, 78.092)
\Line(130.000, 78.092)(130.000, 89.477)
\Line(130.000, 89.477)(140.000, 89.477)
\Line(140.000, 89.477)(140.000,103.169)
\Line(140.000,103.169)(150.000,103.169)
\Line(150.000,103.169)(150.000,118.308)
\Line(150.000,118.308)(160.000,118.308)
\Line(160.000,118.308)(160.000,126.815)
\Line(160.000,126.815)(170.000,126.815)
\Line(170.000,126.815)(170.000,138.969)
\Line(170.000,138.969)(180.000,138.969)
\Line(180.000,138.969)(180.000,148.538)
\Line(180.000,148.538)(190.000,148.538)
\Line(190.000,148.538)(190.000,151.338)
\Line(190.000,151.338)(200.000,151.338)
\Line(200.000,151.338)(200.000,157.846)
\Line(200.000,157.846)(210.000,157.846)
\Line(210.000,157.846)(210.000,155.077)
\Line(210.000,155.077)(220.000,155.077)
\Line(220.000,155.077)(220.000,149.938)
\Line(220.000,149.938)(230.000,149.938)
\Line(230.000,149.938)(230.000,164.462)
\Line(230.000,164.462)(240.000,164.462)
\Line(240.000,164.462)(240.000,158.615)
\Line(240.000,158.615)(250.000,158.615)
\Line(250.000,158.615)(250.000,150.185)
\Line(250.000,150.185)(260.000,150.185)
\Line(260.000,150.185)(260.000,132.200)
\Line(260.000,132.200)(270.000,132.200)
\Line(270.000,132.200)(270.000,103.677)
\Line(270.000,103.677)(280.000,103.677)
\Line(280.000,103.677)(280.000, 87.815)
\Line(280.000, 87.815)(290.000, 87.815)
\Line(290.000, 87.815)(290.000, 49.323)
\Line(290.000, 49.323)(300.000, 49.323)
\Line(300.000, 49.323)(300.000,  0.000)
\DashLine(  0.000,  1.161)( 10.000,  1.161){2}
\DashLine( 10.000,  1.161)( 10.000,  4.196){2}
\DashLine( 10.000,  4.196)( 20.000,  4.196){2}
\DashLine( 20.000,  4.196)( 20.000,  7.816){2}
\DashLine( 20.000,  7.816)( 30.000,  7.816){2}
\DashLine( 30.000,  7.816)( 30.000, 14.024){2}
\DashLine( 30.000, 14.024)( 40.000, 14.024){2}
\DashLine( 40.000, 14.024)( 40.000, 21.760){2}
\DashLine( 40.000, 21.760)( 50.000, 21.760){2}
\DashLine( 50.000, 21.760)( 50.000, 22.760){2}
\DashLine( 50.000, 22.760)( 60.000, 22.760){2}
\DashLine( 60.000, 22.760)( 60.000, 26.380){2}
\DashLine( 60.000, 26.380)( 70.000, 26.380){2}
\DashLine( 70.000, 26.380)( 70.000, 32.040){2}
\DashLine( 70.000, 32.040)( 80.000, 32.040){2}
\DashLine( 80.000, 32.040)( 80.000, 38.376){2}
\DashLine( 80.000, 38.376)( 90.000, 38.376){2}
\DashLine( 90.000, 38.376)( 90.000, 49.320){2}
\DashLine( 90.000, 49.320)(100.000, 49.320){2}
\DashLine(100.000, 49.320)(100.000, 70.360){2}
\DashLine(100.000, 70.360)(110.000, 70.360){2}
\DashLine(110.000, 70.360)(110.000, 74.000){2}
\DashLine(110.000, 74.000)(120.000, 74.000){2}
\DashLine(120.000, 74.000)(120.000, 74.240){2}
\DashLine(120.000, 74.240)(130.000, 74.240){2}
\DashLine(130.000, 74.240)(130.000, 85.560){2}
\DashLine(130.000, 85.560)(140.000, 85.560){2}
\DashLine(140.000, 85.560)(140.000, 93.880){2}
\DashLine(140.000, 93.880)(150.000, 93.880){2}
\DashLine(150.000, 93.880)(150.000,110.680){2}
\DashLine(150.000,110.680)(160.000,110.680){2}
\DashLine(160.000,110.680)(160.000,133.560){2}
\DashLine(160.000,133.560)(170.000,133.560){2}
\DashLine(170.000,133.560)(170.000,139.520){2}
\DashLine(170.000,139.520)(180.000,139.520){2}
\DashLine(180.000,139.520)(180.000,145.040){2}
\DashLine(180.000,145.040)(190.000,145.040){2}
\DashLine(190.000,145.040)(190.000,144.480){2}
\DashLine(190.000,144.480)(200.000,144.480){2}
\DashLine(200.000,144.480)(200.000,153.840){2}
\DashLine(200.000,153.840)(210.000,153.840){2}
\DashLine(210.000,153.840)(210.000,162.560){2}
\DashLine(210.000,162.560)(220.000,162.560){2}
\DashLine(220.000,162.560)(220.000,164.680){2}
\DashLine(220.000,164.680)(230.000,164.680){2}
\DashLine(230.000,164.680)(230.000,163.560){2}
\DashLine(230.000,163.560)(240.000,163.560){2}
\DashLine(240.000,163.560)(240.000,159.360){2}
\DashLine(240.000,159.360)(250.000,159.360){2}
\DashLine(250.000,159.360)(250.000,156.040){2}
\DashLine(250.000,156.040)(260.000,156.040){2}
\DashLine(260.000,156.040)(260.000,139.640){2}
\DashLine(260.000,139.640)(270.000,139.640){2}
\DashLine(270.000,139.640)(270.000,109.960){2}
\DashLine(270.000,109.960)(280.000,109.960){2}
\DashLine(280.000,109.960)(280.000, 88.120){2}
\DashLine(280.000, 88.120)(290.000, 88.120){2}
\DashLine(290.000, 88.120)(290.000, 57.560){2}
\DashLine(290.000, 57.560)(300.000, 57.560){2}
\DashLine(300.000, 57.560)(300.000,  0.000){2}
\DashCurve{( 10.000,  1.500)
( 20.000,  4.544)
( 30.000,  7.968)
( 40.000, 11.536)
( 50.000, 15.492)
( 60.000, 19.528)
( 70.000, 24.208)
( 80.000, 29.360)
( 90.000, 35.560)
(100.000, 42.960)
(110.000, 50.960)
(120.000, 60.480)
(130.000, 69.960)
(140.000, 79.080)
(150.000, 88.680)
(160.000, 98.400)
(170.000,107.480)
(180.000,116.600)
(190.000,125.480)
(200.000,132.320)
(210.000,139.760)
(220.000,145.000)
(230.000,147.880)
(240.000,146.120)
(250.000,140.880)
(260.000,132.600)
(270.000,122.080)
(280.000,105.520)
(290.000, 84.360)(300.000, 58.320)}{0.9} 
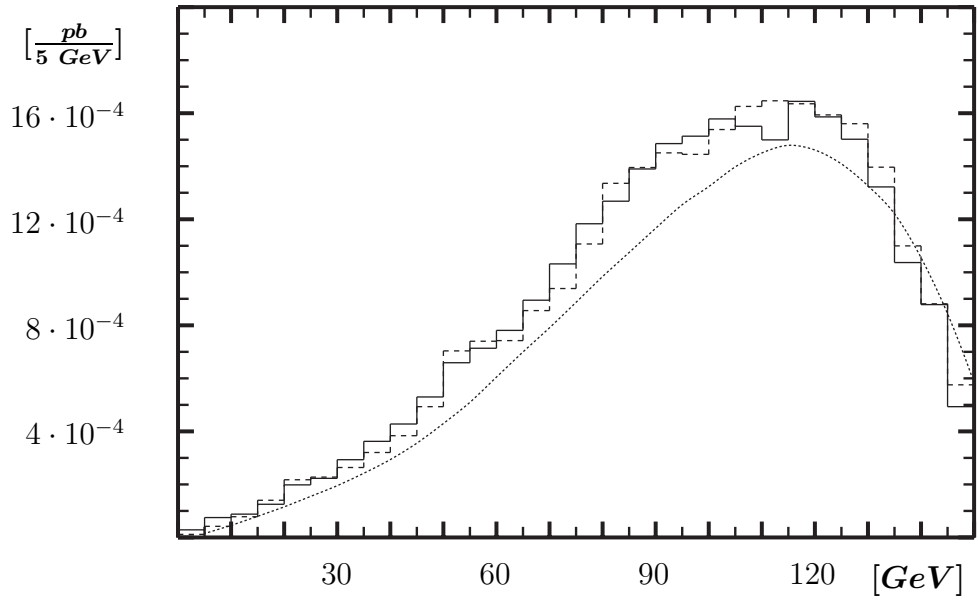
\efig{Figure 5: {\em The histograms are the invariant mass distribution 
of $l^+ + b$ for on-shell (dashed) and off-shell (solid) 
single top production, including final state gluonic corrections.
The dotted line is the off-shell tree level result ($\alpha_s= 0$).}}
 \begin{center} 
  \begin{picture}(300,350)
\SetOffset(70,30)
  \LinAxis(0,0)(200,0)(5,4,5,0,1.5)
  \LinAxis(0,300)(200,300)(5,4,-5,0,1.5)
  \LinAxis(0,0)(0,300)(6,5,-5,0,1.5)
  \LinAxis(200,0)(200,300)(6,5,5,0,1.5)
  \Text(40,-10)[t]{$170$}
  \Text(80,-10)[t]{$174$}
  \Text(120,-10)[t]{$178$}
  \Text(170,-10)[t]{{\boldmath $[GeV]$}}
  \Text(-20,50)[r] {$1.7 \cdot 10^{-3}$}
  \Text(-20,100)[r]{$3.3 \cdot 10^{-3}$}
  \Text(-20,150)[r]{$5.0 \cdot 10^{-3}$}
  \Text(-20,200)[r]{$6.7 \cdot 10^{-3}$}
  \Text(-20,250)[r]{$8.3 \cdot 10^{-3}$}
  \Text(-20,285)[r]{{\boldmath $[\frac{pb}{1~GeV}]$}}
 \Line(  0.000,  6.249)( 10.000,  6.249)
 \Line( 10.000,  6.249)( 10.000,  6.775)
 \Line( 10.000,  6.775)( 20.000,  6.775)
 \Line( 20.000,  6.775)( 20.000, 11.474)
 \Line( 20.000, 11.474)( 30.000, 11.474)
 \Line( 30.000, 11.474)( 30.000, 11.746)
 \Line( 30.000, 11.746)( 40.000, 11.746)
 \Line( 40.000, 11.746)( 40.000, 15.835)
 \Line( 40.000, 15.835)( 50.000, 15.835)
 \Line( 50.000, 15.835)( 50.000, 20.760)
 \Line( 50.000, 20.760)( 60.000, 20.760)
 \Line( 60.000, 20.760)( 60.000, 29.211)
 \Line( 60.000, 29.211)( 70.000, 29.211)
 \Line( 70.000, 29.211)( 70.000, 50.446)
 \Line( 70.000, 50.446)( 80.000, 50.446)
 \Line( 80.000, 50.446)( 80.000, 94.569)
 \Line( 80.000, 94.569)( 90.000, 94.569)
 \Line( 90.000, 94.569)( 90.000,201.692)
 \Line( 90.000,201.692)(100.000,201.692)
 \Line(100.000,201.692)(100.000,172.385)
 \Line(100.000,172.385)(110.000,172.385)
 \Line(110.000,172.385)(110.000, 52.385)
 \Line(110.000, 52.385)(120.000, 52.385)
 \Line(120.000, 52.385)(120.000, 23.543)
 \Line(120.000, 23.543)(130.000, 23.543)
 \Line(130.000, 23.543)(130.000, 12.882)
 \Line(130.000, 12.882)(140.000, 12.882)
 \Line(140.000, 12.882)(140.000,  7.675)
 \Line(140.000,  7.675)(150.000,  7.675)
 \Line(150.000,  7.675)(150.000,  5.557)
 \Line(150.000,  5.557)(160.000,  5.557)
 \Line(160.000,  5.557)(160.000,  3.929)
 \Line(160.000,  3.929)(170.000,  3.929)
 \Line(170.000,  3.929)(170.000,  3.104)
 \Line(170.000,  3.104)(180.000,  3.104)
 \Line(180.000,  3.104)(180.000,  2.271)
 \Line(180.000,  2.271)(190.000,  2.271)
 \Line(190.000,  2.271)(190.000,  1.767)
 \Line(190.000,  1.767)(200.000,  1.767)
 \Line(200.000,  1.767)(200.000,  0.000)
 \DashLine(  0.000,  2.030)( 10.000,  2.030){0.9}
 \DashLine( 10.000,  2.030)( 10.000,  2.543){0.9}
 \DashLine( 10.000,  2.543)( 20.000,  2.543){0.9}
 \DashLine( 20.000,  2.543)( 20.000,  3.218){0.9}
 \DashLine( 20.000,  3.218)( 30.000,  3.218){0.9}
 \DashLine( 30.000,  3.218)( 30.000,  4.315){0.9}
 \DashLine( 30.000,  4.315)( 40.000,  4.315){0.9}
 \DashLine( 40.000,  4.315)( 40.000,  6.132){0.9}
 \DashLine( 40.000,  6.132)( 50.000,  6.132){0.9}
 \DashLine( 50.000,  6.132)( 50.000,  8.926){0.9}
 \DashLine( 50.000,  8.926)( 60.000,  8.926){0.9}
 \DashLine( 60.000,  8.926)( 60.000, 14.736){0.9}
 \DashLine( 60.000, 14.736)( 70.000, 14.736){0.9}
 \DashLine( 70.000, 14.736)( 70.000, 27.792){0.9}
 \DashLine( 70.000, 27.792)( 80.000, 27.792){0.9}
 \DashLine( 80.000, 27.792)( 80.000, 68.172){0.9}
 \DashLine( 80.000, 68.172)( 90.000, 68.172){0.9}
 \DashLine( 90.000, 68.172)( 90.000,202.680){0.9}
 \DashLine( 90.000,202.680)(100.000,202.680){0.9}
 \DashLine(100.000,202.680)(100.000,203.760){0.9}
 \DashLine(100.000,203.760)(110.000,203.760){0.9}
 \DashLine(110.000,203.760)(110.000, 68.844){0.9}
 \DashLine(110.000, 68.844)(120.000, 68.844){0.9}
 \DashLine(120.000, 68.844)(120.000, 28.452){0.9}
 \DashLine(120.000, 28.452)(130.000, 28.452){0.9}
 \DashLine(130.000, 28.452)(130.000, 15.024){0.9}
 \DashLine(130.000, 15.024)(140.000, 15.024){0.9}
 \DashLine(140.000, 15.024)(140.000,  9.332){0.9}
 \DashLine(140.000,  9.332)(150.000,  9.332){0.9}
 \DashLine(150.000,  9.332)(150.000,  6.121){0.9}
 \DashLine(150.000,  6.121)(160.000,  6.121){0.9}
 \DashLine(160.000,  6.121)(160.000,  4.493){0.9}
 \DashLine(160.000,  4.493)(170.000,  4.493){0.9}
 \DashLine(170.000,  4.493)(170.000,  3.468){0.9}
 \DashLine(170.000,  3.468)(180.000,  3.468){0.9}
 \DashLine(180.000,  3.468)(180.000,  2.674){0.9}
 \DashLine(180.000,  2.674)(190.000,  2.674){0.9}
 \DashLine(190.000,  2.674)(190.000,  2.129){0.9}
 \DashLine(190.000,  2.129)(200.000,  2.129){0.9}
 \DashLine(200.000,  2.129)(200.000,  0.000){0.9}
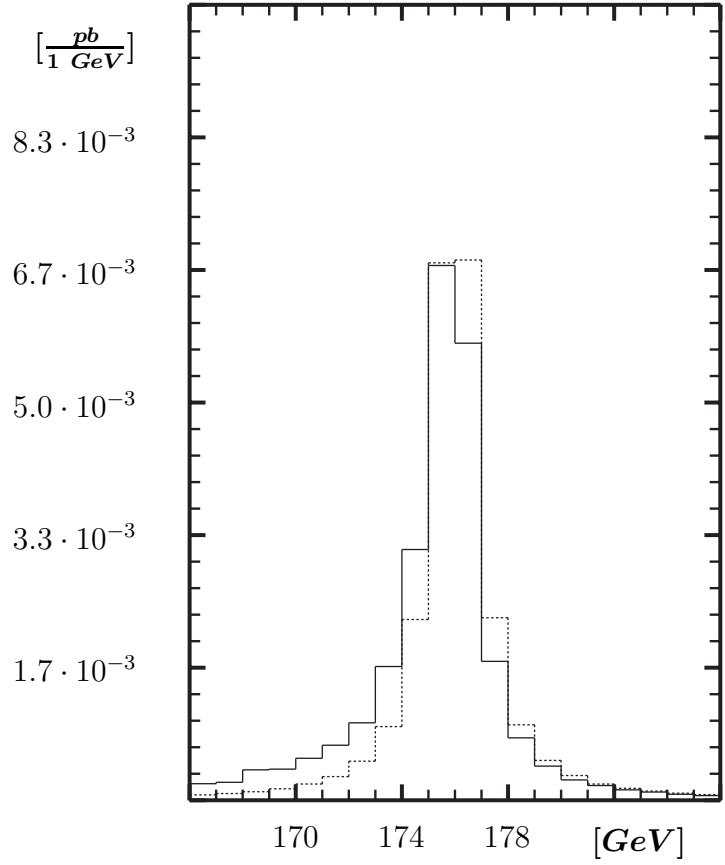
\efig{Figure 6: {\em Invariant mass of $l^+ +  b 
+ \nu_l$ for off-shell single
top production, at the tree level (dotted histogram) and
including final state QCD corrections (solid histogram).}}
\end{document}